\documentclass{iucrjournals}

\usepackage{amsmath,amssymb}
\usepackage{array}
\usepackage{graphicx}
\usepackage{dcolumn}
\usepackage{bm}
\usepackage{booktabs}
\usepackage{comment}
\usepackage{enumitem}
\usepackage[version=4]{mhchem}
\hypersetup{hidelinks}
\usepackage{tabularx}
\usepackage{makecell}

\newcommand{\He}{$^3$He }
\newcommand{\Hensf}{$^3$He-NSF}
\newcolumntype{C}[1]{>{\centering\arraybackslash}p{#1}}
\newcolumntype{Y}{>{\centering\arraybackslash}X}

\title{Polarized neutron diffraction with \textit{ex-situ} $^3$He neutron spin filter and two-dimensional detector for noncollinear incommensurate magnetic order}

\author[a,b]{Shingo~Takahashi}
\author[a]{Yuta~Osawa}
\author[b]{Ryuju~Kobayashi}
\author[c]{Takashi~Ino}
\author[a]{Hiraku~Saito}
\author[b]{Takuro~Kawasaki}
\author[b]{Takayuki~Oku}
\author[b]{Tatsuya~Nakamura}
\author[d]{Yoichi~Ikeda}
\author[d]{Masaki~Fujita}
\author[e]{Daisuke~Koto}
\author[e]{Yusuke~Tokunaga}
\author[e,f]{Taka-hisa~Arima}
\author[a,c,f]{Taro~Nakajima}

\affil[a]{Institute for Solid State Physics, The University of Tokyo, Kashiwa, Chiba 277-8581, Japan}
\affil[b]{J-PARC Center, Japan Atomic Energy Agency, Tokai, Ibaraki 319-1195, Japan}
\affil[c]{Institute of Materials Structure Science, High Energy Accelerator Research Organization, Tsukuba, Ibaraki 305-0801, Japan}
\affil[d]{Institute for Materials Research, Tohoku University, Sendai, Miyagi 980-8577, Japan}
\affil[e]{Department of Advanced Materials Science, The University of Tokyo, Kashiwa, Chiba 277-8561, Japan}
\affil[f]{RIKEN Center for Emergent Matter Science, Wako, Saitama 351-0198, Japan}

\begin{document}
\doublespacing
\maketitle

\begin{synopsis}
Polarized neutron diffraction using an \textit{ex-situ} $^3$He neutron spin filter and a two dimensional detector is demonstrated for a single crystal of the multiferroic material Mn$_3$WO$_6$, which exhibits complex incommensurate magnetic orders.
We established a data reduction procedure that corrects for the effects of imperfect neutron polarization and converts two-dimensional intensity maps into three-dimensional intensity distributions in reciprocal space.
\end{synopsis}

\begin{abstract}
Polarized neutron scattering has been widely used for studying magnetic orders in condensed matter.
A typical setup for this technique is a triple-axis spectrometer with a ferromagnetic single-crystal monochromator and analyzer, which is suited for point-by-point measurements within the horizontal scattering plane.
However, this setup cannot fully meet the demands of recent studies on complex magnetic orders with emergent cross-correlated phenomena.
For instance, multiferroics, magnetic skyrmions, and other topologically nontrivial magnetic orders tend to have incommensurate magnetic modulation vectors running along various directions in reciprocal space, which cannot be captured by measurements restricted to the horizontal scattering plane.
To overcome these limitations, we have developed an experimental setup combining a \He neutron spin filter (\Hensf) and a two-dimensional (2D) detector, which enables us to analyze the polarization of neutrons scattered away from the horizontal plane.
We also establish a data reduction procedure to convert the 2D intensity maps into three-dimensional intensity distributions in reciprocal space, correcting for the effect of imperfect \He polarization.
This method is applied to single-crystal diffraction measurements on the multiferroic material \ce{Mn3WO6}, which exhibits complex incommensurate magnetic orders.
We have successfully observed polarization-dependent intensities of out-of-plane incommensurate magnetic reflections, demonstrating the effectiveness of the present method for analyzing complex magnetic structures.
\end{abstract}


\section{Introduction}
\begin{figure}[htbp]
    \centering
    \includegraphics[width=\linewidth]{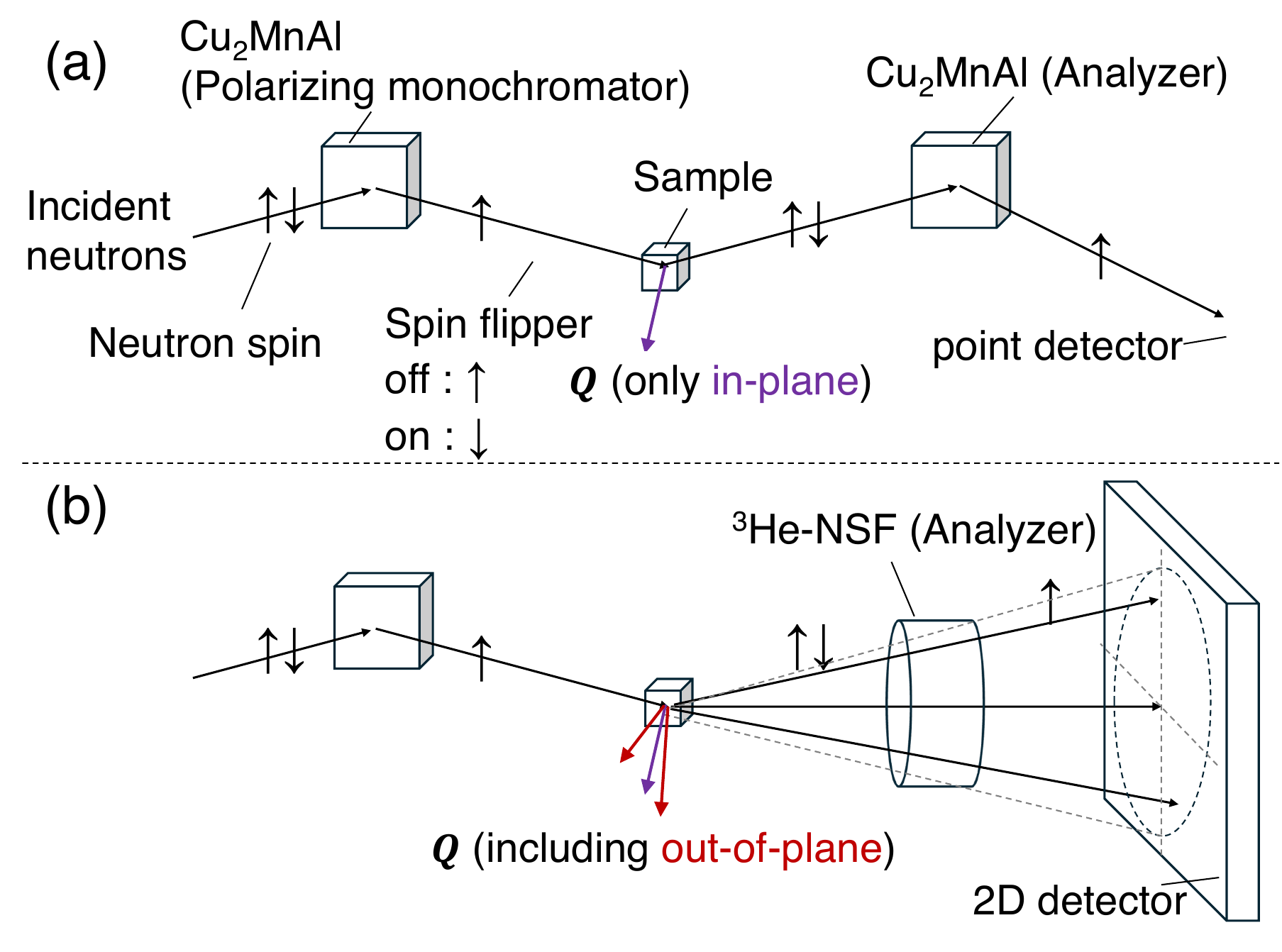}
    \caption{Schematic illustrations comparing neutron polarization analysis techniques.
(a) Conventional setup using Heusler crystals for both the polarizer and the analyzer, where only in-plane scattered neutrons are detected by a point detector.
(b) Polarization analysis setup combining a \Hensf~and a 2D detector, enabling detection of both in-plane and out-of-plane scattered neutrons.}
    \label{fig:introduction}
\end{figure}

Polarized neutron scattering is a powerful technique for investigating magnetic order and excitations in condensed matter.
This capability arises from the magnetic moment of the neutron, which directly interacts with magnetic moments of electrons in the material through the magnetic dipole-dipole interaction.
By analyzing the spin states of neutrons before and after scattering, one can separate magnetic scattering from nuclear scattering and determine the directions of magnetic moments in the sample.
To perform these measurements, devices for controlling and probing the neutron polarization are required.
To date, three techniques have been widely used for polarized neutron scattering.
The first is a single crystal of ferromagnetic material, such as a \ce{Cu2MnAl} Heusler compound.
By exploiting the nuclear-magnetic interference effect, which gives rise to different Bragg reflectivities for two neutron spin states, a polarized monochromatic neutron beam can be obtained \cite{squires1996introduction}.
The second is a polarizing supermirror, which utilizes the difference in critical angle between neutrons with different spin states \cite{mezei1976novel}.
The third is a polarizing filter, which utilize the strong spin dependence of the neutron absorption cross section of \He nuclei (\He neutron spin filter, \Hensf) \cite{gentile2017optically}.
Among these techniques, the ferromagnetic single-crystal monochromators are often used to polarize incident neutrons and analyze scattered neutrons in triple-axis spectrometers \cite{kulda_in20_2002,faulhaber_polarized_2010,fernandez-baca_triple-axis_2008,takeda_double_2008,nakajima_polarized_2024}, as shown in Fig.~\ref{fig:introduction}(a).
While this type of instrument has been widely used for studying magnetic structures, accessible reflections are often restricted to the horizontal scattering plane.

In recent years, increasing attention has been paid to materials hosting complex incommensurate magnetic orders, such as spin-driven multiferroics, magnetic skyrmions and related topological magnetic orders~\cite{muhlbauer_skyrmion_2009,tokura_multiferroics_2014,kanazawa_direct_2020,fujishiro_topological_2019,kurumaji_electronic_2025}.
In these systems, the magnetic moments are often arranged in noncollinear or noncoplanar configurations, which can give rise to emergent cross-correlated phenomena.
To study these systems, it is necessary to measure magnetic satellite reflections corresponding to magnetic modulation wavevectors ($q$-vectors), which often run along various directions in reciprocal space.
Consequently, three-dimensional exploration of the reciprocal space with a two-dimensional (2D) detector is of great importance.
To realize polarization analysis in the measurements with a 2D detector, polarization devices with a large solid-angle coverage, such as \Hensf, are required.

The combination of a \Hensf~and a 2D detector has been implemented in several small-angle neutron scattering (SANS) instruments and applied to studies of complex magnetic structures \cite{singh_transition_2023,takagi_multiple-q_2018}.
However, their accessible reciprocal space region is generally limited to relatively low-$Q$.
To extend the accessible $Q$ range for polarization analysis to middle-$Q$ and high-$Q$ regions, it would be necessary to place a \Hensf~and a 2D detector at various scattering angles, as illustrated in Fig.~\ref{fig:introduction}(b).
Related instrumental concepts have already been reported at several neutron facilities, including POLI at MLZ~\cite{thoma_polarized_2018} and WOMBAT at ANSTO~\cite{lee_polarised_2016}.
Nevertheless, quantitative experimental demonstrations of polarization analysis using this type of setup remain limited.
The \Hensf~has also been used in the two-axis diffractometer D1B at ILL~\cite{heil_large_2002}. %
This instrument has a one-dimensional (1D) array of detectors, which can cover a wide range of the scattering angle in a horizontal plane, and is dedicated to powder diffraction measurements. %
As for inelastic scattering instruments, the \He-NSFs have been used in MACS at NIST~\cite{fu_wide_2011,ye_wide_2013,chen_recent_2016} and  LET at ISIS~\cite{nilsen_polarisation_2017,cassella_polarization_2019}; the former and latter have 1D and 2D detectors, respectively. %
It should be mentioned that another neutron inelastic spectrometer HYSPEC in ORNL also has polarization analysis capability using not \Hensf~ but radial supermirror analyzers and 2D detectors~\cite{zaliznyak_polarized_2017}.
These extensive developments of polarized neutron instruments indicate a growing demand for polarized neutron measurements in condensed matter physics. %

In the present study, we performed polarized neutron diffraction measurements combining a \Hensf~and a 2D detector on the 5G beamport at JRR-3.
This paper mainly describes the experimental setup, the measurement procedures, and corrections for the effects of imperfect neutron polarization.
We also performed polarized neutron diffraction on a single crystal of \ce{Mn3WO6}, which exhibits incommensurate magnetic orders.
This compound has a crystal structure belonging to the space group $R\bar{3}$ and displays two magnetically ordered phases at low temperatures \cite{maignan_spin-induced_2020}.
As we describe in the following sections, we focus on the data measured at 23 K, which corresponds to the higher-temperature magnetic phase.
We performed polarization analysis for magnetic reflections that were slightly shifted from the horizontal plane.
The details of the magnetic structure will be reported elsewhere (Osawa et al., work in preparation).

\section{Experimental details}

\subsection{Setup}
The 5G beamport at JRR-3 is normally operated as the POlarized Neutron Triple-Axis spectrometer (PONTA), which has a longitudinal polarization analysis option with conventional Heusler polarizing monochromator and analyzer~\cite{nakajima_polarized_2024}.
In the present experiment, the Heusler analyzer and the point detector were replaced with a \Hensf~and a 2D detector.
The Heusler polarizing monochromator provided a polarized incident neutron beam with an energy of 14.7 meV.
The higher-order reflections from the monochromator were suppressed by a pyrolytic graphite filter placed between the monochromator and the sample.

\begin{figure}[htbp]
    \centering
    \includegraphics[width=\linewidth]{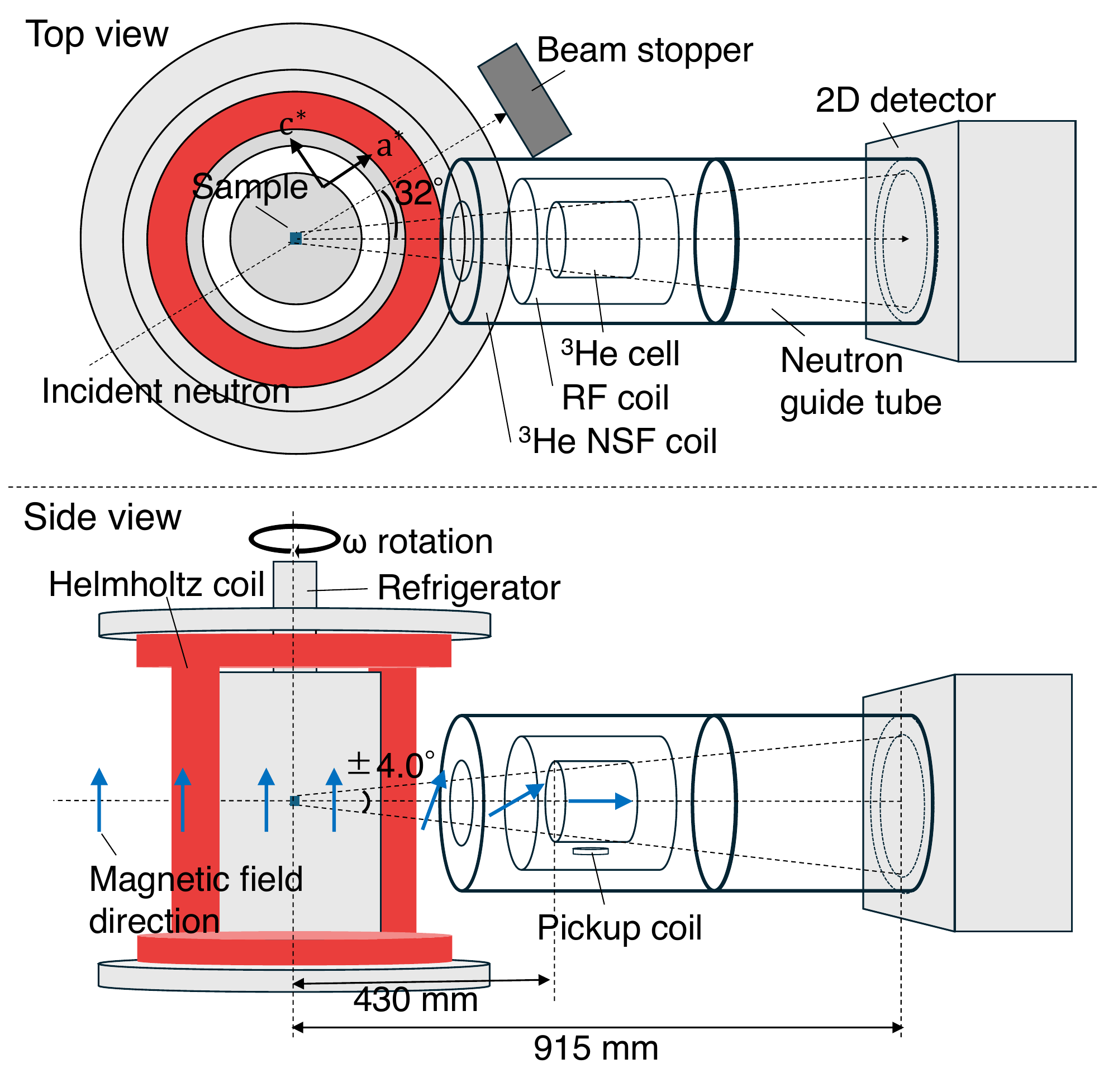}
    \caption{
    Schematic illustration of the experimental setup on the 5G beamport.
    Scattered neutrons from the sample pass through the \Hensf~coil and the neutron guide tube before reaching the 2D detector.
    The sample is mounted in a cryostat whose rotation angle is denoted by $\omega$.
    The magnetic field direction induced by the Helmholtz coil is vertical, whereas the field within the \Hensf~is oriented along the beam axis.
    The neutron spins are adiabatically rotated to follow the field direction.
    A radio-frequency (RF) coil is used to flip the \He polarization, and an NMR pickup coil monitors the \He polarization after each spin flip.
    }
    \label{fig:Setup}
\end{figure}

Figure~\ref{fig:Setup} shows an overview of the experimental setup between the sample and the detector.
Scattered neutrons were detected using a scintillation-type 2D detector employing wavelength-shifting fiber readout~\cite{nakamura_large-area_2012, noauthor_compact_nodate}.
The active area was $256 \times 256$ mm$^2$, consisting of 64 $\times$ 64 channels with a channel pitch of 4 mm.
To access the target reflections, the detector was positioned at an angle of $32^\circ$ from the incident direction throughout the experiment.
The sample-to-detector distance was 915 mm.
A \Hensf~was installed 430 mm downstream of the sample.
The cylindrical \He cell provided an angular acceptance of $\pm4.0^\circ$, which corresponded to a circular region with a radius of 76 mm (approximately 19 channels) centered on the detector.
Polarization analysis of the scattered neutrons was possible only within this region.
To reduce the background, a \ce{B4C}-shielded neutron guide tube was installed between the \Hensf~and the detector, and \ce{B4C} shielding was also attached around the \Hensf~coil.
The non-spin-flip and spin-flip intensities of the scattered neutrons were obtained by flipping the \He nuclear polarization using the adiabatic fast passage nuclear magnetic resonance (AFP NMR) \cite{abragam1961principles}.
Therefore, the neutron spin flipper for the incident beam was not used.
A Helmholtz coil was used to maintain the neutron spin polarization along the neutron flight path by applying a vertical magnetic field.
The magnetic field in the \Hensf~was oriented parallel to the beam axis, and the spins of neutrons scattered from the sample adiabatically followed the field direction, as illustrated in Fig.~\ref{fig:Setup}.
The magnetic field strength generated by both the \Hensf~coil and the Helmholtz coil was approximately 1 mT.

A single-crystal sample of \ce{Mn3WO6} with a volume of 8 mm$^3$ was mounted on an aluminum holder and installed in a $^4$He closed-cycle refrigerator.
This compound has a rhombohedral crystal structure belonging to the space group of $R\bar{3}$. %
The lattice constants were reported to be $a = b$ = 8.8905(6) \AA~and $c$ = 10.4734(8) \AA~in the hexagonal setting~\cite{maignan_spin-induced_2020}.
The $(H,0,L)$ plane was aligned with the horizontal scattering plane.
The sample rotation angle $\omega$ about the vertical axis was adjusted for each target reflection to satisfy the corresponding Bragg condition.

\subsection{\He neutron spin filter}
\label{sec:hensf}
\begin{table}[t]

\caption{Specifications of \He cells.}
\label{tab:He_cell}
    \centering
    \small
    \begin{tabular*}{\linewidth}{@{\extracolsep{\fill}}lccccc@{}}
        \toprule
        Name
        & $D \times L$ (mm)
        & Volume (cm$^3$)
        & \He density (amg)
        & N$_2$ pressure (Torr)
        & K/Rb (mol)\\
        \midrule
        Kenji
        & $72 \times 67$
        & 140
        & 3.2
        & 97
        & 26\\
        Shimpachi
        & $72 \times 65$
        & 146
        & 3.1
        & 100
        & 37\\
        \bottomrule
    \end{tabular*}

\end{table}
In this section, we describe the \Hensf~employed in the present experiment in detail.
Neutrons with spins antiparallel to the \He nuclear spins are strongly absorbed, while those with spins parallel to the \He nuclear spins are transmitted.
Thus, neutrons are polarized by passing through the polarized \He gas, and neutron polarization and transmission depend on the \He polarization.
There are two major methods for polarizing \He nuclei, one is spin-exchange optical pumping (SEOP) \cite{bouchiat_nuclear_1960}, and the other is metastability-exchange optical pumping (MEOP) \cite{colegrove1963polarization}.
In the present study, we employed the SEOP method.
In the SEOP method, generally, a glass cell is filled with $^3$He, \ce{N2}, and a small amount of alkali metal.
The cell is irradiated with circularly polarized light resonant with an optical transition of the alkali metal.
The alkali-metal electrons undergo repeated excitation and de-excitation, with \ce{N2} serving as a quenching gas to suppress radiative decay, resulting in electron spin polarization.
The polarization of the alkali metal is then transferred to the \He nuclei via spin-exchange collisions, thereby achieving \He nuclear polarization.

The SEOP system used in this study was a modified version of the system developed for the polarized neutron spectrometer POLANO, installed at BL23 in MLF of J-PARC \cite{ino_development_2017}.
Both the \Hensf~coil for polarization analysis shown in Fig.~\ref{fig:Setup} and that for the SEOP system were equipped with an AFP NMR system \cite{ino_compact_2012}.
The AFP NMR system consisted of a radio-frequency (RF) coil for flipping the \He polarization and a pickup coil located at the bottom of the \He cell for monitoring the \He polarization.
Two cylindrical \He glass cells, named Kenji and Shimpachi, made of aluminosilicate glass (GE180) with nearly identical dimensions, were used in this experiment.
Both cells were filled with $^3$He, \ce{N2}, and a mixture of Rb and K for hybrid SEOP, which provided a higher pumping efficiency than \ce{Rb}-only SEOP \cite{babcock_hybrid_2003,chen_spin-exchange_2007}.
The details of the cell specifications are described in Table~\ref{tab:He_cell}.

The \Hensf~was operated in an \textit{ex-situ} configuration, in which the \He cell was polarized in the SEOP system and then transported to the beamline for neutron measurements.
A darkroom housing the SEOP laser system was constructed in the 5G beamport experimental area.
After the \He polarization reached saturation, the \He cell was transported to the \Hensf~coil on the beamline.
Neutron measurements were performed while the \He polarization relaxed, resulting in decreases in both the neutron polarization and transmission.
Since the \He polarization decreased with time, two cells were used alternately.
While one cell was used for polarized-neutron measurements, the other was polarized in the SEOP system.
The cells were exchanged every two or three days.

\section{Data reduction procedure including polarization correction}
\label{sec:datareduction}
\begin{figure}[htbp]
    \centering
    \includegraphics[width=\linewidth]{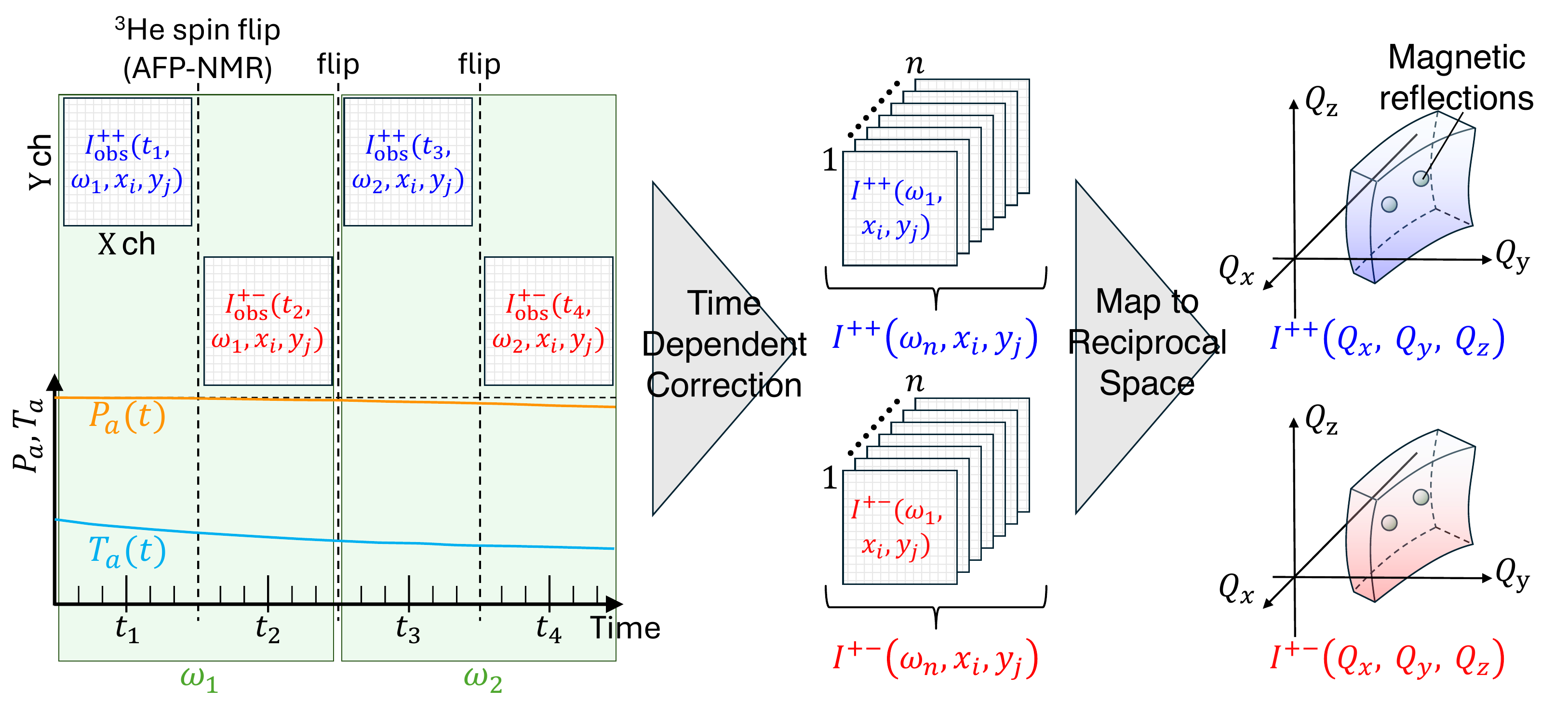}
    \caption{
    Schematic diagram of the data-reduction procedure for polarization analysis using an \textit{ex-situ} \Hensf~and a 2D detector.
    Raw 2D detector data measured at different sample rotation angles $\omega$ and times $t$ are corrected for the time-dependent neutron polarization $P_a(t)$ and transmission $T_a(t)$ of the \Hensf.
    The detector coordinates (X,Y) and sample rotation angle $\omega$ are converted to $Q_x,Q_y,Q_z,$ and then transformed to $H,K,L$ using the UB matrix.
    The corrected three-dimensional reciprocal-space data can then be sliced along an arbitrary plane.
    }
    \label{fig:procedure}
\end{figure}

Figure~\ref{fig:procedure} shows the procedure for data acquisition, reduction, and polarization correction in the present study.
At each sample rotation angle $\omega=\omega_n$, we measured a pair of 2D intensity maps corresponding to non-spin-flip and spin-flip scattering by flipping the \He polarization.
The measurement time for each intensity map was 20 min for magnetic reflections and 5 min for nuclear reflections.
We denote the observed intensities measured with the \He nuclear polarization parallel and antiparallel to the neutron polarization produced by the Heusler polarizer as $I^{++}_\mathrm{obs}$ and $I^{+-}_\mathrm{obs}$, respectively.
All intensity maps were normalized by the measurement time.
The measurements of $I^{++}_\mathrm{obs}$ and $I^{+-}_\mathrm{obs}$ were repeated while varying the sample rotation angle $\omega$.

To account for the time dependence of the \textit{ex-situ} \Hensf, the neutron polarization and transmission during each measurement were approximated by their values at the midpoint of the measurement.
As shown in the left panel of Fig.~\ref{fig:procedure}, the midpoint times of two successive measurements at $\omega=\omega_n$ are denoted by $t_{2n-1}$ and $t_{2n}$.

In a full polarization analysis experiment, one would measure four scattering intensities, $I^{++}$, $I^{+-}$, $I^{-+}$, and $I^{--}$, corresponding to the four possible combinations of the polarizer and analyzer spin states~\cite{babcock_discussion_2017}.
Among these scattering intensities, $I^{++}$ and $I^{--}$ are the non-spin-flip channels, and $I^{+-}$ and $I^{-+}$ are the spin-flip channels.
In the present experiment, the nuclear and magnetic reflections were separated, and chiral magnetic scattering was negligible because the neutron polarization and the scattering vector were nearly perpendicular to each other~\cite{blume_polarization_1963}.
We therefore assumed $I^{++}=I^{--}$ and $I^{+-}=I^{-+}$, allowing the polarization analysis to be performed using only the two measured channels.

Using these relations and the midpoint times defined above, the observed intensities at detector pixel $(x_i,y_j)$ are expressed as

    \begin{align}
        I^{++}_\mathrm{obs}(\omega_n,t_{2n-1}, x_i,y_j) &= {T_pT_a(t_{2n-1})}\biggl[\left(1+ P_pP_a(t_{2n-1})\right)I^{++}(\omega_n, x_i,y_j) \notag\\
        &+\left(1- P_pP_a(t_{2n-1})\right)I^{+-}(\omega_n, x_i,y_j)\biggr]+\mathrm{BG}(x_i,y_j),
        \label{eq:int_obs_pp}\\
        I^{+-}_\mathrm{obs}(\omega_n,t_{2n}, x_i,y_j) &= {T_pT_a(t_{2n})}\biggl[\left(1- P_pP_a(t_{2n})\right)I^{++}(\omega_n, x_i,y_j) \notag\\
        &+\left(1+ P_pP_a(t_{2n})\right)I^{+-}(\omega_n, x_i,y_j)\biggr]+\mathrm{BG}(x_i,y_j),
        \label{eq:int_obs_pm}
    \end{align}

where $I^{++}(\omega_n,x_i,y_j)$ and $I^{+-}(\omega_n,x_i,y_j)$ are the ideal non-spin-flip and spin-flip scattering intensities, respectively.
$T_p$ and $T_a(t)$ denote the neutron reflectance of the Heusler polarizer and the transmission of the \Hensf, respectively.
$P_p$ and $P_a(t)$ are the neutron polarizations produced by the Heusler polarizer and \Hensf, respectively.
Note that the transmission and the polarization are defined for an unpolarized incident neutron beam.
For the Heusler polarizer, $T_p$ is defined as the ratio of the Bragg-reflected intensity to the incident unpolarized neutron intensity.
Although $P_a(t)$ is generally defined as a signed quantity ranging from $-1$ to $+1$, we define it as its magnitude, $0 \leq P_a(t) \leq 1$.
Instead, the parallel and antiparallel orientations of the \He polarization relative to the incident neutron polarization are represented by the signs in the factors $\pm P_pP_a(t)$ in Eqs.~\ref{eq:int_obs_pp} and \ref{eq:int_obs_pm}.
The quantities $T_p$ and $P_p$ were treated as time-independent, whereas $T_a(t)$ and $P_a(t)$ were time-dependent because of the relaxation of the \He nuclear polarization.
$\mathrm{BG}(x_i,y_j)$ is the background intensity at the detector pixel $(x_i, y_j)$.
In the present experiment, the background signals were mainly due to imperfect shielding of the detector and the neutron flight path.
Thus, we assumed that the background signals were independent of time, \He polarization, and the sample rotation angle $\omega$. %
We measured the $\mathrm{BG}(x_i,y_j)$ by setting the $\omega$ angle away from any Bragg reflections from the sample.

By solving Eqs.~\ref{eq:int_obs_pp} and \ref{eq:int_obs_pm} for $I^{++}$ and $I^{+-}$, we obtained the equations of the polarization correction for the dataset measured at $\omega=\omega_n$ as follows:

    \begin{align}
        I^{++}(\omega_n,x_i,y_j) =&
        \frac{1+P_pP_a(t_{2n})}
        {2T_pT_a(t_{2n-1})P_p\left[P_a(t_{2n-1})+P_a(t_{2n})\right]}
        \left[I^{++}_\mathrm{obs}(\omega_n,t_{2n-1},x_i,y_j)-\mathrm{BG}(x_i,y_j)\right]
        \notag\\
        &-
        \frac{1-P_pP_a(t_{2n-1})}
        {2T_pT_a(t_{2n})P_p\left[P_a(t_{2n-1})+P_a(t_{2n})\right]}
        \left[I^{+-}_\mathrm{obs}(\omega_n,t_{2n},x_i,y_j)-\mathrm{BG}(x_i,y_j)\right],
        \\
        I^{+-}(\omega_n,x_i,y_j) =&
        -
        \frac{1-P_pP_a(t_{2n})}
        {2T_pT_a(t_{2n-1})P_p\left[P_a(t_{2n-1})+P_a(t_{2n})\right]}
        \left[I^{++}_\mathrm{obs}(\omega_n,t_{2n-1},x_i,y_j)-\mathrm{BG}(x_i,y_j)\right]
        \notag\\
        &+
        \frac{1+P_pP_a(t_{2n-1})}
        {2T_pT_a(t_{2n})P_p\left[P_a(t_{2n-1})+P_a(t_{2n})\right]}
        \left[I^{+-}_\mathrm{obs}(\omega_n,t_{2n},x_i,y_j)-\mathrm{BG}(x_i,y_j)\right].
        \label{eq:int_corr}
    \end{align}

As we show later, we also experimentally determined the neutron polarizations produced by the polarizer and the analyzer, $P_p$ and $P_a(t)$, and the transmission of the analyzer, $T_a(t)$, including the temporal decay of the \He polarization.
Since $T_p$ is time-independent, it was instead treated as a constant scale factor.
Substituting all the parameters into the equations above, we corrected the effect of imperfect polarization for all the pixels.

Finally, we converted the detector coordinates and sample rotation angle $(x_i, y_j, \omega_n)$ into reciprocal-space coordinates $(Q_x,Q_y,Q_z)$.
The resulting data were further transformed into $(H,K,L)$ coordinates using the UB matrix, allowing the intensity distribution to be sliced along an arbitrary plane in reciprocal space.

\section{Results}
\subsection{Evaluation of \He neutron spin filter from nuclear scattering}
\label{sec:He-3}

The \He polarization was evaluated by measuring a nuclear Bragg peak.
Specifically, we selected the $\bar{1}02$ reflection, which has a similar scattering angle to that of the magnetic reflections of interest.
The neutron transmission $T_a(t)$, the neutron polarization $P_a(t)$, and the \He polarization $P_\mathrm{He}(t)$ are given by the following equations \cite{coulter_neutron_1990}:
\begin{align}
    T_a(t) &= T_\mathrm{cell}\exp(-\mathcal{O})\cosh(\mathcal{O}P_\mathrm{He}(t)),
    \label{eq:transmission_3He}\\
    P_a(t) &= \tanh(\mathcal{O}P_\mathrm{He}(t)),
    \label{eq:pol_n3He}\\
    P_\mathrm{He}(t) &= P_\mathrm{He}(0) \exp(-t/\tau).
    \label{eq:pol_3He}
\end{align}
Here, $T_\mathrm{cell}$ is the transmission of the glass walls of the \He cell.
$\mathcal{O}$ is the opacity of the \He cell, defined as $\mathcal{O} = \rho d \sigma$.
$\rho$ is the number density of \He, and $d$ is the neutron path length through the \He cell, which depends on the positions of the detector pixels.
$\sigma$ is the neutron absorption cross section of unpolarized \He, which depends on the incident neutron energy.
According to the literature, $\sigma$ for the incident neutron energy of $14.7~\textrm{meV}$ is $6997~\textrm{barn}$ \cite{passell_measurement_1966}.
$P_{\mathrm{He}}(0)$ is the initial polarization at $t=0$, and $\tau$ is the relaxation time.

First, the areal density $\rho d$ of each cell was determined as follows.
Before polarizing \He by SEOP, we measured the nuclear peak through the unpolarized \He cell with $\omega$ fixed at the Bragg condition.
From the intensity map, we extracted a box-integrated intensity near the peak.
After subtracting the background signals, we obtained the integrated intensity of the Bragg reflection, $\mathcal{I}_\mathrm{obs}^\mathrm{unpol}$.
By using Eqs.~\ref{eq:int_obs_pp}, \ref{eq:transmission_3He}, and \ref{eq:pol_3He}, and assuming that the \He polarization is zero ($P_\mathrm{He}(t)=0$), $\mathcal{I}_\mathrm{obs}^\mathrm{unpol}$ is expressed as follows:
\begin{equation}
    \mathcal{I}^\mathrm{unpol}_\mathrm{obs}
    =\sum_{(i,j)\in R}
    T_p
    T_\mathrm{cell}
    \exp(-\mathcal{O})
    I^{++}(x_i,y_j)
    .
    \label{eq:unpol}
\end{equation}
where $R$ denotes the box-integrated region.
Note that the nuclear Bragg reflection contains only the non-spin-flip scattering intensity, and thus we can impose $I^{+-} = 0$ in Eq.~\ref{eq:int_obs_pp}.
The same measurement was performed using an empty cell with the same dimensions as the \He glass cell, yielding $\mathcal{I}^\mathrm{empty}_\mathrm{obs}$.
Here, the transmission term $T_\mathrm{cell}\exp(-\mathcal{O})$ is replaced by $T_\mathrm{cell}$.
Then, taking the ratio $\mathcal{I}^\mathrm{unpol}_\mathrm{obs}/\mathcal{I}^\mathrm{empty}_\mathrm{obs}$, the $\rho d$ values of the Kenji and Shimpachi cells were determined to be $18.5 \pm 0.5~\mathrm{amg\,cm}$ and $18.8 \pm 0.4~\mathrm{amg\,cm}$, respectively.
Note that the amagat is a non-SI unit of the volumetric number density of an ideal gas at 0 $^{\circ}$C (273.15 K) and 1 atm (101.325 kPa), and one amagat is equivalent to $2.69\times 10^{19} /\textrm{cm}^3$.
The $\rho d$ values determined here correspond to the path lengths through the \He gas at the pixel positions of the detected nuclear peaks, and the path length $d$ varies with pixel position.
However, the solid angle covered by the cell was sufficiently small (up to $4^\circ$) that the variation in $d$ could be neglected.
Therefore, these $\rho d$ values were used for all pixels included in the present data analysis.
$T_{\mathrm{cell}}$ was determined to be $0.91 \pm 0.04$ by comparing the nuclear peak intensities measured without the cell with the empty cell.
This value was not directly used in the polarization analysis.

\begin{figure}[htbp]
    \centering
    \includegraphics[width=\linewidth]{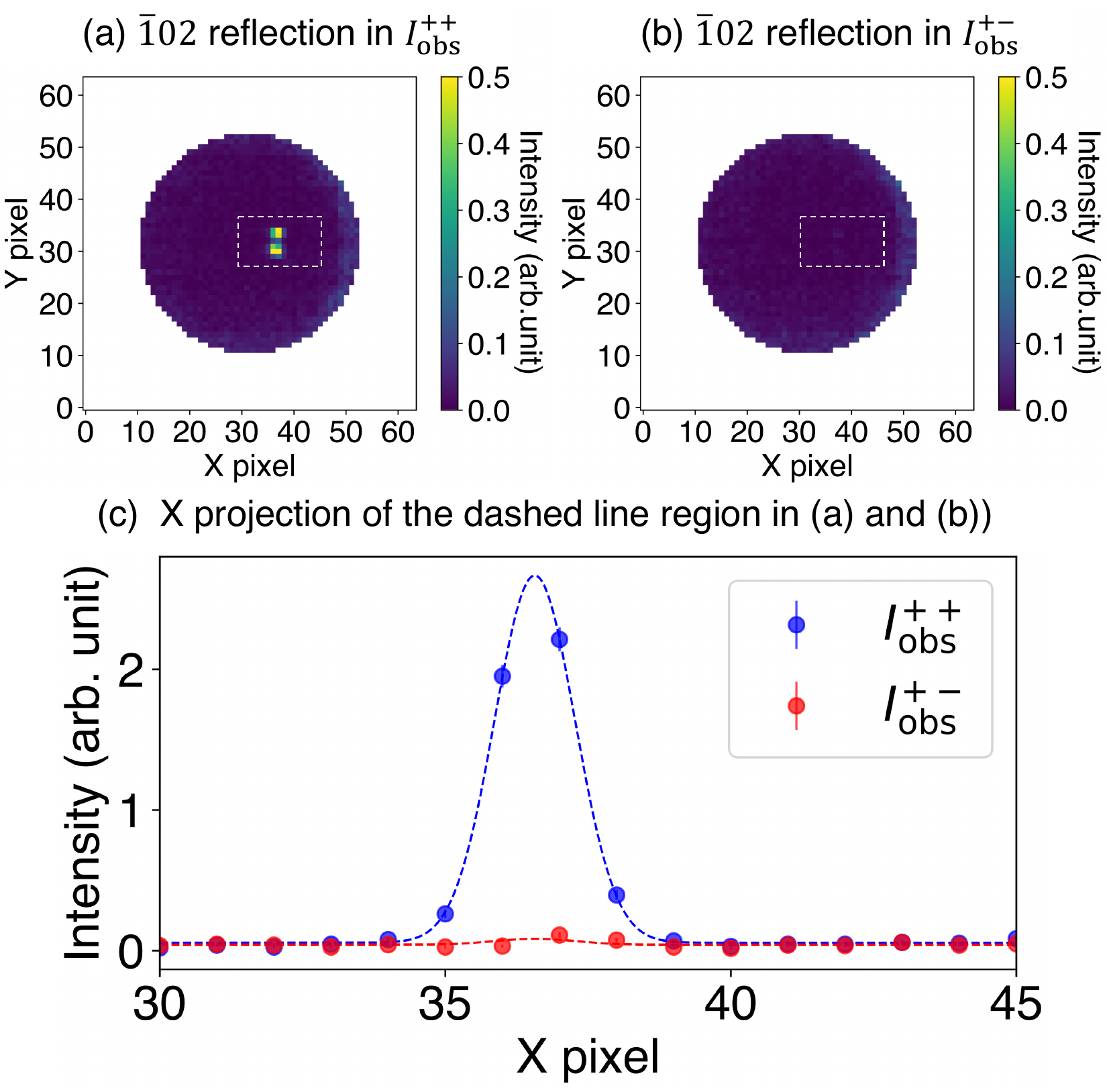}
    \caption{(a) Observed non-spin-flip ($I^{++}_\mathrm{obs}$) and (b) spin-flip ($I^{+-}_\mathrm{obs}$) intensity maps, showing only the analyzing region.
    (c) Line profile along the $X$ direction, obtained by integrating the intensity within the rectangular regions indicated by the dashed lines in (a) and (b).
    }
    \label{fig:rawdata}
\end{figure}

Second, we determined $P_\mathrm{He}(0)$ in Eq.~\ref{eq:pol_3He} by measuring the nuclear reflection with the polarized \He cell.
Immediately after moving the \He cell from the SEOP system to the beamline, we measured the intensities of the reflection with the \He polarization parallel and antiparallel to the incident neutron polarization.
The integrated intensities were obtained following the same procedure as that used for the measurement with the unpolarized \He cell.
These are referred to as $\mathcal{I}^\mathrm{++}_\mathrm{obs}(t=0)$ and $\mathcal{I}^\mathrm{+-}_\mathrm{obs}(t=0)$, respectively.
Since these measurement times were much shorter than the \He relaxation time, the \He polarizations for the two measurements were assumed to be equal.
As shown in Fig.~\ref{fig:rawdata}, we clearly observed the nuclear Bragg peak in the non-spin-flip channel, while the intensity in the spin-flip channel was extremely weak.
Note that the peak in Fig.~\ref{fig:rawdata}(a) shows a slight splitting due to the mosaicity of the single-crystal sample.
From Eqs.~\ref{eq:int_obs_pp}, \ref{eq:int_obs_pm}, \ref{eq:transmission_3He}, and \ref{eq:unpol}, the \He polarization $P_\mathrm{He}(t)$ at $t=0$ satisfies the following relationship:
\begin{align}
    \frac{\mathcal{I}^{++}_\mathrm{obs}(t=0)+\mathcal{I}^{+-}_\mathrm{obs}(t=0)}{\mathcal{I}^\mathrm{unpol}_\mathrm{obs}} = 2\cosh(\mathcal{O}P_\mathrm{He}(0)).
    \label{eq:transmission_ratio}
\end{align}
We measured the nuclear Bragg reflection to estimate $P_\mathrm{He}(0)$ each time the cell was changed.
For instance, the initial \He polarization of Shimpachi cell was determined to be $P_\mathrm{He}(0)=0.81\pm0.05$.

\begin{figure}[htbp]
    \centering
    \includegraphics[width=\linewidth]{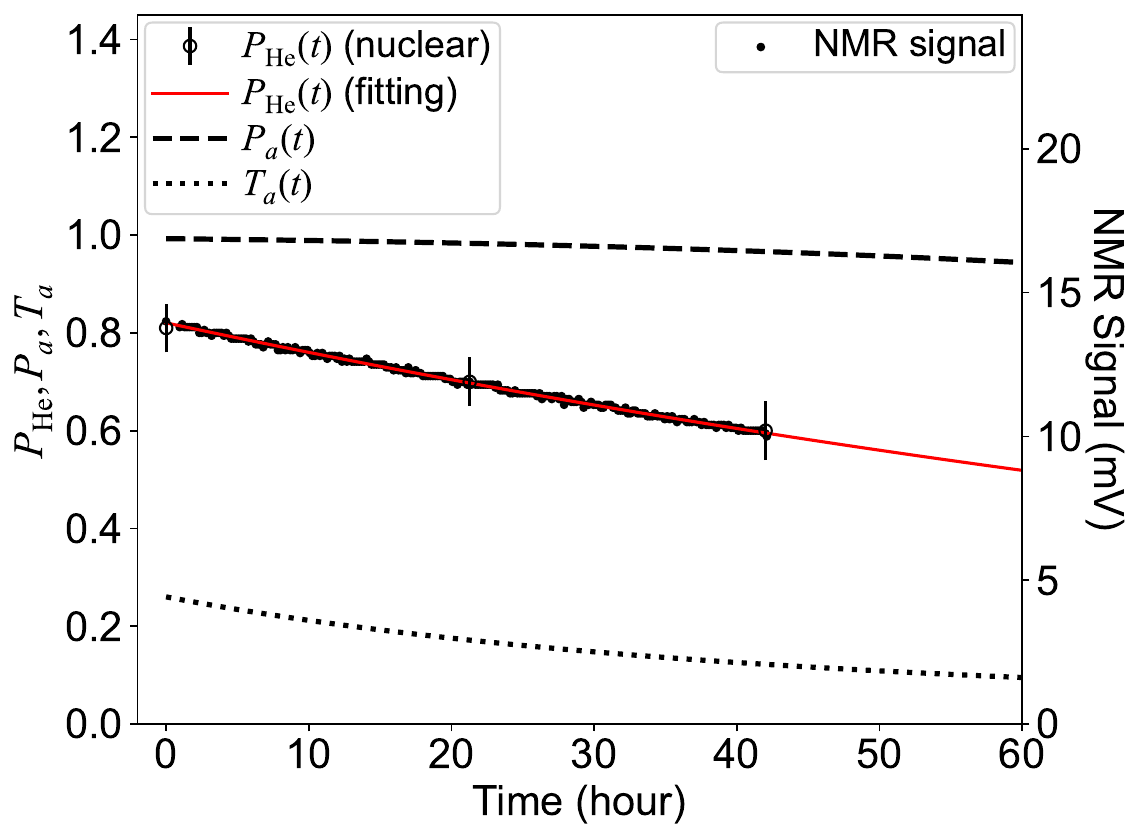}
    \caption{
        Time dependence of the \He polarization, neutron polarization, and neutron transmission.
        The open circles show the \He polarization determined from the intensity of the $\bar{1}02$ nuclear reflection, while the closed circles show the NMR signal.
        The NMR signal was normalized to the initial \He polarization determined from the reflection.
        The red line represents the fitting result using Eq.~\ref{eq:pol_3He}.
        The dashed and dotted lines indicate the calculations for $P_a(t)$ and $T_a(t)$, respectively, obtained from $P_\mathrm{He}(t)$.
        }
        \label{fig:pol_trans_decay}
\end{figure}

We then measured the relaxation of the \He polarization by tracking the NMR signal of the \Hensf, which is proportional to the \He polarization.
Throughout the neutron diffraction measurement, the NMR signal was recorded each time the \He polarization was flipped.
Figure~\ref{fig:pol_trans_decay} shows the time dependence of the NMR signal of Shimpachi cell.
By normalizing the NMR data to the $P_\mathrm{He}(0)$ determined by the nuclear reflection, we obtained the \He polarization as a function of time.
We fitted Eq.~\ref{eq:pol_3He} to the normalized NMR data and obtained $P_\mathrm{He}$ as shown by the red solid curve in Fig.~\ref{fig:pol_trans_decay}.
The relaxation times $\tau$ for Shimpachi and Kenji cells were determined to be $\tau = 131.1 \pm 0.7$ and $\tau = 128.4 \pm 0.3$ hours, respectively.
These relaxation times were much longer than the measurement time for each 2D intensity map (5 min for nuclear reflections and 20 min for magnetic reflections), indicating that the decay of the \He polarization during each measurement was negligible.
We also measured the nuclear Bragg reflection several times during the experiment, confirming that the values of the \He polarization deduced from the measurements of the nuclear reflection and the NMR signal showed good agreement. %
The time evolutions of $T_a(t)$ and $P_a(t)$ were also obtained using Eqs.~\ref{eq:transmission_3He}, \ref{eq:pol_n3He}, and \ref{eq:pol_3He}, as shown in Fig.~\ref{fig:pol_trans_decay}.

The neutron polarization produced by the Heusler polarizer, $P_p$, was also determined from the integrated intensities of the nuclear reflection.
From Eqs.~\ref{eq:int_obs_pp} and \ref{eq:int_obs_pm}, the asymmetry between $\mathcal{I}^{++}_\mathrm{obs}$ and $\mathcal{I}^{+-}_\mathrm{obs}$ after subtracting the background $\mathrm{BG}(x_i, y_j)$ from each is given by
\begin{equation}
    \frac{\mathcal{I}^{++}_\mathrm{obs}-\mathcal{I}^{+-}_\mathrm{obs}}{\mathcal{I}^{++}_\mathrm{obs}+\mathcal{I}^{+-}_\mathrm{obs}} = P_pP_a(t).
\end{equation}
Since $P_a(t)$ has already been determined from the \He polarization, $P_p$ can be obtained from this equation.
The values of $P_p$ deduced from the measurements of the nuclear reflection were averaged, yielding $P_p = 0.86 \pm 0.06$.
Note that $P_p$ includes the effects of environmental background and neutron depolarization in the guide field.

\begin{figure}[htbp]
    \centering
    \includegraphics[width=\linewidth]{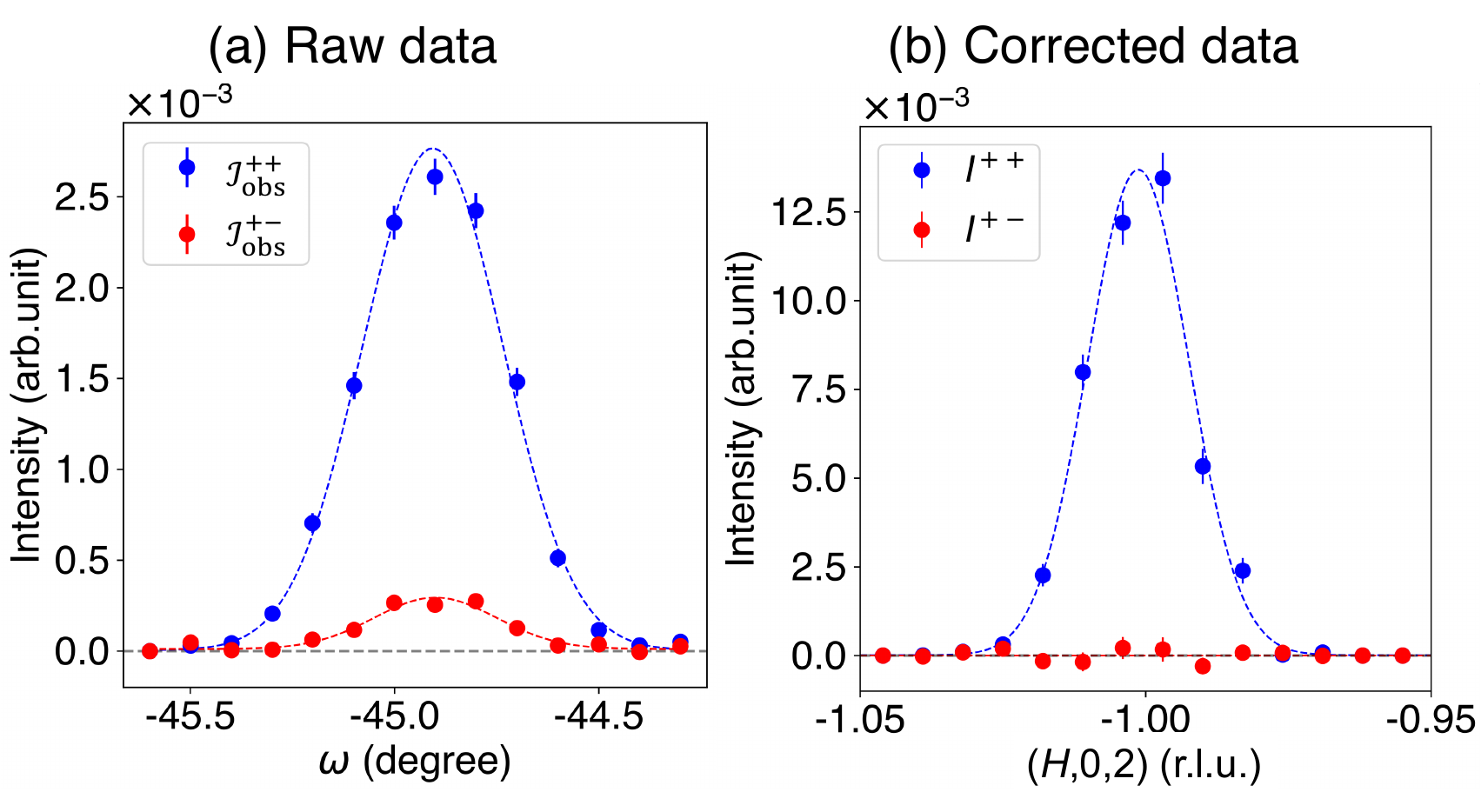}
    \caption{
    (a) Raw non-spin-flip and spin-flip intensities, $\mathcal{I}^{++}_\mathrm{obs}$ (red) and $\mathcal{I}^{+-}_\mathrm{obs}$ (blue), obtained by integrating the peak intensity on the 2D detector while scanning the sample rotation angle $\omega$.
    (b) Corrected non-spin-flip and spin-flip scattering intensities, $I^{++}(H,K,L)$ (red) and $I^{+-}(H,K,L)$ (blue) in reciprocal space.
    The data were obtained by slicing along the $(H,0,2)$ line in the three-dimensional reciprocal space and integrating over a thickness of $\pm$ 0.025 r.l.u. along the $(K/2,-K,0)$ direction, which is perpendicular to the horizontal $(H0L)$ scattering plane, and over $\pm$ 0.03 r.l.u. along the $L$ direction.
    }
    \label{fig:polcorrection}
\end{figure}

To check the validity of the polarization correction, we measured the intensity distributions near the $\bar{1}02$ reflection while varying the $\omega$ angle.
Figure~\ref{fig:polcorrection}(a) shows the observed box-integrated intensities, $\mathcal{I}^{++}_\mathrm{obs}(\omega_n, t_{2n-1})$ and $\mathcal{I}^{+-}_\mathrm{obs}(\omega_n, t_{2n})$.
The polarization correction was not applied to these data.
Non-zero $I^{+-}_\mathrm{obs}$ was observed owing to the imperfect neutron polarization of the polarizer and analyzer. %
Then, we applied the polarization correction to all the observed intensity maps and converted them into an intensity distribution in reciprocal space.
Figure~\ref{fig:polcorrection}(b) shows the profiles of the corrected non-spin-flip and spin-flip intensities, ${I}^{++}$ and ${I}^{+-}$, along the $(H,0,2)$ line, which were sliced from the three-dimensional data.
This result indicates that the polarization correction was successfully performed.
Details of the polarization correction are described in Appendix~\ref{appendix}.



\subsection{Polarization analysis of magnetic scattering}
\begin{figure}[htbp]
    \centering
    \includegraphics[width=\linewidth]{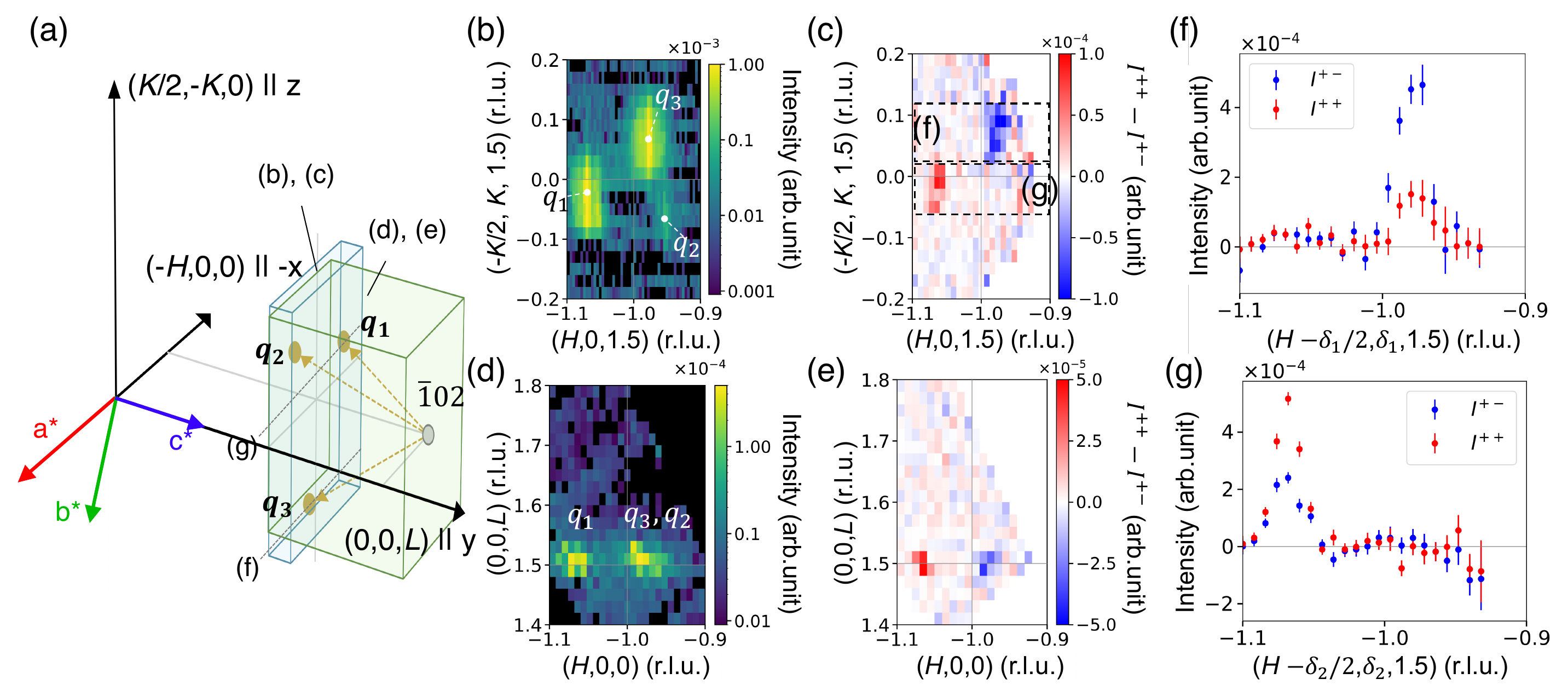}
    \caption{
(a) Schematic illustration of the reciprocal-space region measured in the present experiment.
The translucent volumes labeled (b), (c) and (d), (e) indicate the regions integrated along the \(L\) direction and the \((-K/2,K,0)\) direction, respectively.
The dotted lines along the \(-H\) direction through \(\boldsymbol{q}_3\) and \(\boldsymbol{q}_1\) indicate the line cuts shown in panels (f) and (g), respectively.
(b), (d) Unpolarized neutron intensity maps in the \((H,K,1.5)\) and \((H,0,L)\) planes, respectively.
(c), (e) Polarization-analysis maps obtained from the difference between the corrected non-spin-flip and corrected spin-flip intensities, \(I^{++}-I^{+-}\).
The intensity maps in the \((H,K,1.5)\) plane shown in (b) and (c) were obtained from the three-dimensional reciprocal space data by integrating over $1.47 \le L \le 1.52$ r.l.u.
Similarly, the intensity maps in the $(H,0,L)$ plane shown in (d) and (e) were obtained by integrating along the \((-1/2,1,0)\) direction over the range of $\pm0.1$ r.l.u.
(f), (g) Line profiles of $I^{++}$ and $I^{+-}$ through $\boldsymbol{q}_3$ and $\boldsymbol{q}_1$, respectively, obtained by projecting the intensities within the dotted boxes in (c) onto the horizontal axis.
Here, the coordinate along the \((-1/2,1,0)\) direction is parameterized by $\delta$, such that the horizontal axis is given by $(H-\delta/2,\delta,1.5)$.
The projected ranges were $0.015\le\delta_1\le0.095$ for (f) and $-0.07\le\delta_2\le0.02$ for (g).
}
    \label{fig:measured_Qregion}
\end{figure}

We applied the polarization analysis with the \Hensf~and the 2D detector to the study of magnetic structures of \ce{Mn3WO6}.
As mentioned in the introduction, this compound displays two antiferromagnetically ordered phases.
In this paper, we focus on the high-temperature phase and show the data measured at 23 K.
First, we determined the $q$-vector of the magnetic order by unpolarized neutron diffraction measurements with the 2D detector.
The horizontal scattering plane was set to the $(H,0,L)$ plane with the c axis perpendicular to the incident neutron beam.
We measured intensity maps by scanning the $\omega$ angle from $-52.9^\circ$ to $-46.0^\circ$ in steps of $0.1^\circ$ and converted the resulting data into reciprocal-space coordinates.
Figure~\ref{fig:measured_Qregion}(d) shows the intensity distribution projected onto the $(H,0,L)$ plane.
We observed two incommensurate reflections near $(-1,0,1.5)$.
To investigate the position of the reflections in more detail, we plotted the intensity distribution on the $(H,K,1.5)$ plane, which is perpendicular to the horizontal plane, as shown in Fig.~\ref{fig:measured_Qregion}(b).
This reveals three incommensurate peaks surrounding $(-1,0,1.5)$.
One of the reflections (labeled $q_1$ in Fig.~\ref{fig:measured_Qregion}(b)) can be indexed using the $q$-vector $\boldsymbol{q}_1 = (0.059, 0.019, 0.5)$.
The other two reflections can be indexed using $\boldsymbol{q}_2$ and $\boldsymbol{q}_3$, which are obtained by threefold rotations of $\boldsymbol{q}_1$ about the $c^*$ axis.
This relationship is consistent with the threefold rotational symmetry of the crystal structure about the $c$ axis.
Note that the unequal distribution of the magnetic intensities would be due to the orientations of the Fourier-transformed magnetic moments with respect to the direction of the scattering vector.
We emphasize here that $\boldsymbol{q}_1$, $\boldsymbol{q}_2$, and $\boldsymbol{q}_3$ cannot lie within a single 2D plane passing through the reciprocal-space origin as shown in Fig.~\ref{fig:measured_Qregion}(a).
This highlights the importance of three-dimensional reciprocal-space measurements for elucidating such complex incommensurate magnetic structures. %

We thus measured the spin-flip and non-spin-flip intensity distributions around $(-1,0,1.5)$ using the \Hensf.
Figures~\ref{fig:measured_Qregion}(c) and \ref{fig:measured_Qregion}(e) show the intensity difference $I^{++}-I^{+-}$ mapped onto the $(H,K,1.5)$ and $(H,0,L)$ planes, respectively.
We clearly observed that the difference in intensity takes positive and negative values for the reflections $\boldsymbol{q}_1$ and $\boldsymbol{q}_3$, respectively.
To compare the spin-dependent intensities more quantitatively, line profiles along the $H$ direction were obtained by projecting the intensities within the dashed-box regions corresponding to $\boldsymbol{q}_3$ and $\boldsymbol{q}_1$ in Fig.~\ref{fig:measured_Qregion}(c).
The resulting line profiles are shown in Figs.~\ref{fig:measured_Qregion}(f) and \ref{fig:measured_Qregion}(g), respectively.
At the reflection corresponding to $\boldsymbol{q}_3$, we found that the magnetic scattering intensity was dominated by spin-flip scattering, while at $\boldsymbol{q}_1$, the non-spin-flip scattering was more prominent.
The asymmetry of the line profiles, defined as $(I^{++}-I^{+-})/(I^{++}+I^{+-})$, was calculated from the integrated intensities, and the results are summarized in Table~\ref{tab:asymmetry}.

\begin{figure}[htbp]
    \centering
    \includegraphics[width=\linewidth]{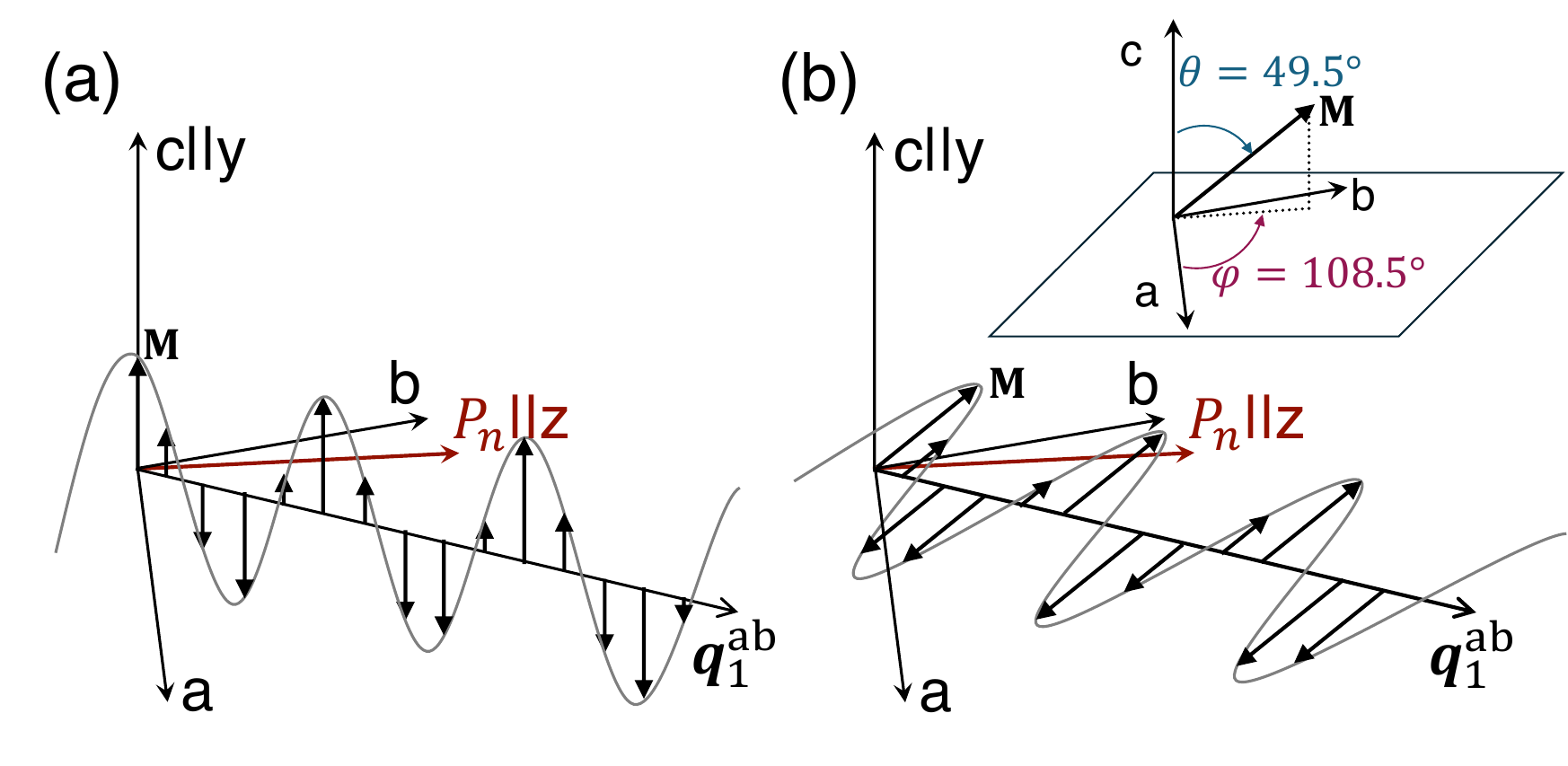}
\caption{Schematic illustrations of the sinusoidal magnetic structures.
(a) the $c$-axis-parallel sinusoidal model and (b) the oblique sinusoidal model.
The $\boldsymbol{q}_1^\mathrm{ab}$ vector is the projection of the magnetic propagation vector $\boldsymbol{q}_1$ onto the $ab$ plane.
The $P_n$ vector is the neutron polarization direction, which is parallel to the $z$ axis in Fig.~\ref{fig:measured_Qregion}(a).
Inset of the panel (b) shows the definition of the polar angle $\theta$ and azimuthal angle $\phi$.
In the oblique sinusoidal model, the magnetic moment direction $\mathbf{M}$ is specified by $\theta = 49.5^\circ$ and $\phi = 108.5^\circ$.
}
    \label{fig:qvector}
\end{figure}

We compared the measured asymmetries with those calculated for different magnetic moment directions.
Since the neutron polarization is parallel to the $z$ axis defined in Fig.~\ref{fig:measured_Qregion}(a), the spin-flip scattering arises from the Fourier-transformed magnetization components in the $xy$ plane, which contains the $a$ and $c$ axes.
If the magnetic moments have only a $c$-axis component, as illustrated in Fig.~\ref{fig:qvector}(a), the calculated asymmetry is $-0.99$ for both reflections.
In this case, both reflections would appear predominantly in the spin-flip channel, which is inconsistent with the experimental results.
We therefore considered an oblique sinusoidal model, as illustrated in Fig.~\ref{fig:qvector}(b).
The moment direction is specified by the polar angle $\theta$, measured from the $c$ axis, and the azimuthal angle $\phi$, measured from the $a$ axis.
A least-squares analysis was performed to determine $\theta$ and $\phi$, yielding $\theta=49.5^\circ$ and $\phi=108.5^\circ$.
A more detailed structural analysis, including data from other regions of reciprocal space, will be reported elsewhere (Osawa et al., work in preparation).
These results demonstrate that the combination of a \Hensf~and a 2D detector enables polarization analysis of magnetic reflections distributed in three-dimensional reciprocal space.
\begin{table}[t]
  \centering
  \caption{Measured and calculated asymmetries for the
  $(-1,0,2)-\boldsymbol{q}_1$ and $(-1,0,2)-\boldsymbol{q}_3$ reflections.
  The asymmetry is defined as $(I^{++}-I^{+-})/(I^{++}+I^{+-})$.
  The experimental values were obtained by integrating the line profiles
  shown in Figs.~7(f) and 7(g). The $c$-axis-parallel and oblique sinusoidal
  models of the magnetic moment $\mathbf{M}$ are illustrated in
  Figs.~8(a) and 8(b), respectively.
  }
  \label{tab:asymmetry}
  \small
  \begin{tabularx}{\columnwidth}{Ycc}
    \toprule
    & $(-1,0,2)-\boldsymbol{q}_1$
    & $(-1,0,2)-\boldsymbol{q}_3$ \\
    \midrule
    Experiment values
    & $0.30 \pm 0.04$
    & $-0.49 \pm 0.10$ \\

    \makecell[c]{$\mathbf{M}\parallel c$ sinusoidal model
    $(\theta=0^\circ,\ \phi=0^\circ)$}
    & $-0.99$
    & $-0.99$ \\

    \makecell[c]{Oblique sinusoidal model
    $(\theta=49.5^\circ,\ \phi=108.5^\circ)$}
    & $0.30$
    & $-0.50$ \\
    \bottomrule
  \end{tabularx}
\end{table}

\section{Summary and Outlook}
In this work, we demonstrated polarized neutron diffraction measurements combining the \Hensf~and the 2D detector to explore three-dimensional reciprocal space.
We established the correction procedure for the time-dependent neutron polarization and transmission using a nuclear Bragg reflection and the NMR signal, and confirmed its validity.
The polarization analysis for the incommensurate magnetic reflections of \ce{Mn3WO6} was successfully performed, highlighting its importance for magnetic reflections outside the horizontal plane.
Although the present \textit{ex-situ} method requires time-dependent correction, it is possible to upgrade it to an \textit{in-situ} SEOP method.
In fact, a compact \textit{in-situ} \Hensf~system has been recently developed at J-PARC~\cite{takahashi_development_2025}.
This implementation is expected to maintain high neutron polarization and transmission throughout the experiment, thereby facilitating polarization analysis of complex magnetic structures over a wider region of reciprocal space and for a broader range of materials.

\begin{acknowledgements}
We thank Mr. Manabu Ohkawara for his kind support in operating the SEOP laser system in the 5G beamport experimental area, and Dr. Takuya Okudaira and Dr. Shusuke Takada for their constructive discussions.
We are also grateful to Dr. Earl Babcock and Dr. Jonathan White for their valuable comments on the manuscript.
The polarized neutron scattering experiment on the 5G beamport at JRR-3 was carried out under Proposal No. 25401.
\end{acknowledgements}

\begin{funding}
This work was supported by JSPS KAKENHI Grant Number 25KJ0050.
The construction of the NSF system was supported by JSPS KAKENHI
Grant Number JP21H04987, with partial support from JP24K21682.
\end{funding}

\appendix
\section{Polarization correction with an \textit{ex-situ} \He neutron spin filter}
\label{appendix}
\begin{figure}[htbp]
    \centering
    \includegraphics[width=\linewidth]{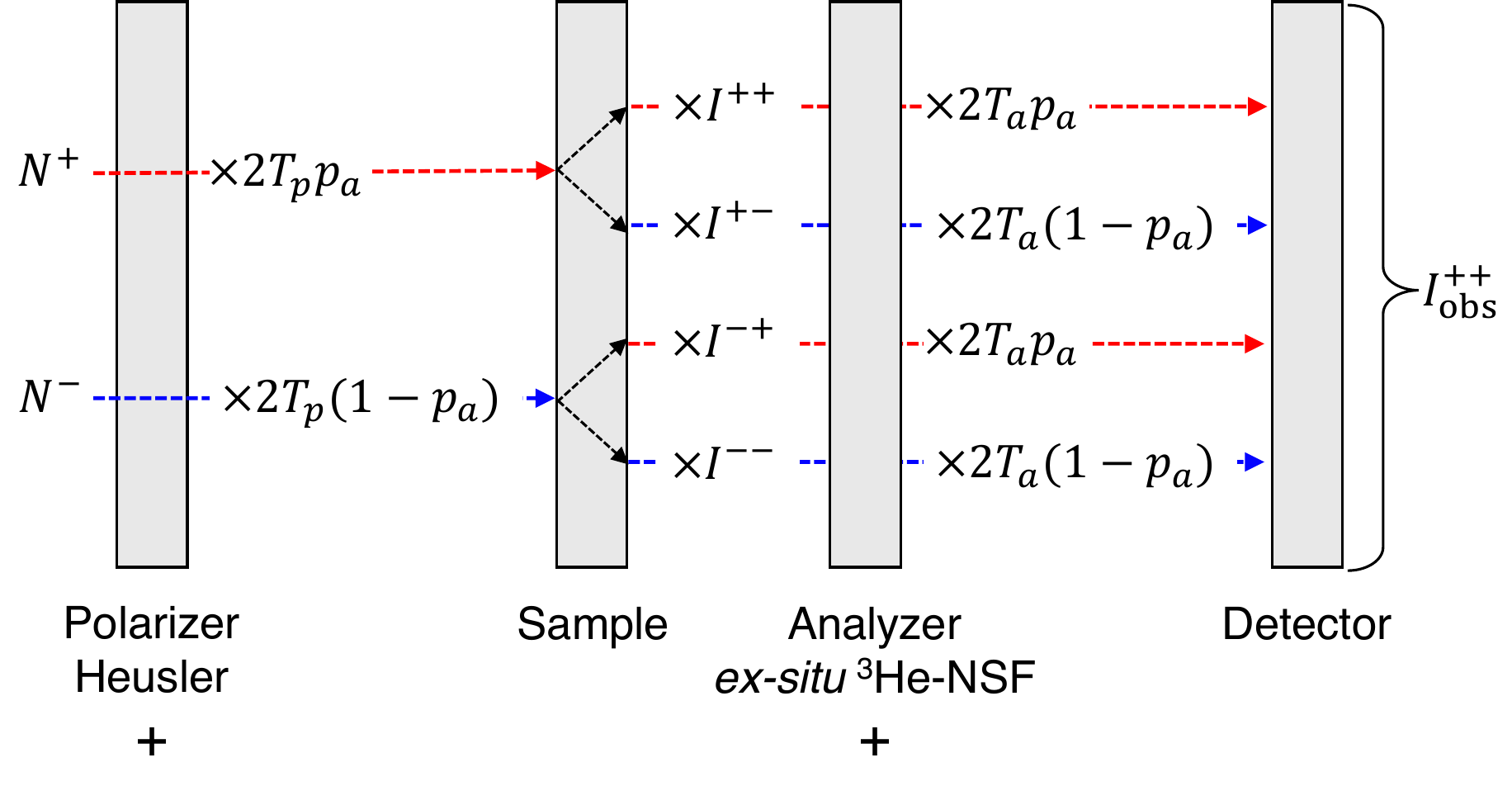}
    \caption{Schematic diagram of the spin-dependent neutron intensities in the present experimental setup when the Heusler polarizer and the \Hensf~are both set to select up-spin ($+$) neutrons.
    The red and blue lines indicate the paths of the up neutrons ($N^+$) and down neutrons ($N^-$), respectively.
    The observed intensity, $I^{++}_\mathrm{obs}$, is obtained as the sum of the contributions from the four spin-scattering channels.}
    \label{fig:appendix}
\end{figure}

In this section, the polarization correction in the present experimental setup is described.
This correction is based on the polarization matrix reported by A. Wildes \cite{wildes_polarizer-analyzer_1999,wildes_scientific_2006}.
A count obtained when both the polarizer and the analyzer are set to the spin-up state, $I^{++}_\mathrm{obs}$, is expressed as the sum of the contributions from the four scattering channels:
\begin{align}
    I^{++}_\mathrm{obs} &= N^+ 2T_p p_p I^{++} 2T_a p_a \notag \\
    &+ N^+ 2T_p p_p I^{+-} 2T_a (1-p_a) \notag\\
    &+ N^- 2T_p (1-p_p) I^{-+} 2T_a p_a \notag \\
    &+ N^- 2T_p (1-p_p) I^{--} 2T_a (1-p_a).
    \label{eq:Ipp}
\end{align}
Here, $N^+$ and $N^-$ denote the numbers of up and down neutrons incident on the polarizer from the neutron source, respectively.
$T_p$ and $T_a$ are the unpolarized neutron reflectance and the transmissions of the polarizer and analyzer, respectively.
The polarization efficiencies of the polarizer and analyzer are defined as $p_p = (1+P_p)/2$ and $p_a = (1+P_a)/2$, respectively, where $P_p$ and $P_a$ are the neutron polarization magnitudes produced by the polarizer and analyzer.
These parameters take values in the ranges $0 \leq T_p,\, T_a,\, P_p,\,P_a  \leq 1$, and $0.5 \leq p_p,\, p_a \leq 1$.
$I^{++}$, $I^{+-}$, $I^{-+}$, and $I^{--}$ are the scattering intensities for the corresponding spin states.
The derivation of this equation is schematically illustrated in Fig.~\ref{fig:appendix}.
In practice, this equation also includes a background term, which is ignored here for simplicity.

In the present experimental setup, only the $I^{++}_\mathrm{obs}$ and $I^{+-}_\mathrm{obs}$ configurations can be measured.
If the \Hensf~analyzer is switched to the spin-down state, the factors for the up and down neutrons, $2T_a p_a$ and $2T_a(1-p_a)$, are interchanged.
Here, we define the total number of incident neutrons as $N_0 = N^+ + N^-$.
Assuming an unpolarized incident beam coming from the neutron source, we have $N^+ = N^- = N_0/2$.
$N_0$ is usually normalized by the counting time or monitor counts.
As also noted in Sec.~\ref{sec:datareduction}, the sample is an antiferromagnet and its nuclear and magnetic reflections are separated from each other.
Moreover, the chiral magnetic scattering, which depends on the sign of the incident neutron polarization direction, can be neglected because the neutron polarization is nearly perpendicular to the scattering vector.
With these conditions, $I^{++} = I^{--}$ and $I^{+-} = I^{-+}$ hold, allowing Eq.~\ref{eq:Ipp} to be simplified into the following matrix form in terms of the non-spin-flip and spin-flip intensities:
\begin{equation}
    \begin{bmatrix}
        I^{++}_\mathrm{obs} \\
        I^{+-}_\mathrm{obs}
    \end{bmatrix}
    =
    2T_pT_a
    \begin{bmatrix}
        p_p & 1-p_p \\
        1-p_p & p_p \\
    \end{bmatrix}
    \begin{bmatrix}
        p_a&1-p_a \\
        1-p_a&p_a \\
    \end{bmatrix}
    \begin{bmatrix}
        I^{++} \\
        I^{+-}
    \end{bmatrix}.
    \label{eq:Ipp_Ipm}
\end{equation}

Since the \He polarization of the \textit{ex-situ} \Hensf~analyzer decays over time, both the neutron polarization efficiency $p_a$ and the neutron transmission $T_a$ become time-dependent.
Accordingly, by treating Eq.~\ref{eq:Ipp_Ipm} as a time-dependent matrix equation and taking its inverse, the following expression is obtained:

\begin{equation}
\begin{aligned}
\begin{bmatrix}
    I^{++} \\
    I^{+-}
\end{bmatrix}
&=
\frac{1}{2T_pP_p\left(P_a(t_1)+P_a(t_2)\right)}
\begin{bmatrix}
    1+P_pP_a(t_2) & -1+P_pP_a(t_1) \\
    -1+P_pP_a(t_2) & 1+P_pP_a(t_1)
\end{bmatrix}
\\
&\quad{}\times
\begin{bmatrix}
    1/T_a(t_1) & 0 \\
    0 & 1/T_a(t_2)
\end{bmatrix}
\begin{bmatrix}
    I^{++}_{\mathrm{obs}}(t_1) \\
    I^{+-}_{\mathrm{obs}}(t_2)
\end{bmatrix}.
\end{aligned}
\label{eq:Sigma}
\end{equation}

where $p_p$ and $p_a$ are converted to $P_p$ and $P_a$ using the relations $P_p = 2p_p-1$ and $P_a = 2p_a-1$.
$T_a(t)$, $P_p$, and $P_a(t)$ are determined experimentally, as described in Sec.~\ref{sec:He-3}.
In contrast, $T_p$ of the Heusler polarizer does not depend on time and is treated as a single constant scale factor.
This time-dependent correction was applied to each detector pixel, allowing data collected throughout the entire measurement period to be used for the polarization analysis.

\bibliography{ref}

\end{document}